# Anomalous propagation of luminescence through bulk n-InP

Serge Luryi, Oleg G. Semyonov, Arsen Subashiev, and Zhichao Chen

*State University of New York at Stony Brook, Dept. of Electrical Engineering, Stony Brook, NY 11794*

**Abstract:** Implementation of a semiconductor as a scintillator with a lattice-matched surface photo-diode for radiation detection requires efficient luminescence collection. Low and heavily doped bulk n-InP has been studied to optimize luminescence transmission via photon recycling.

The key issue in implementing a semiconductor as luminescent material is how to make the material transparent to its own luminescence, so that photons generated deep inside the semiconductor slab could reach its surface without tangible attenuation. However, direct-gap semiconductors are opaque at wavelengths corresponding to their emission spectrum. One of the possibilities is to make a semiconductor relatively transparent by doping it heavily with donor impurities, so as to introduce the Burstein shift between the emission and the absorption spectra [1]. Another approach [2] is based on the high radiative efficiency of high-quality direct-gap semiconductors, such as InP and GaAs. In these materials, an act of interband absorption does not finish off a scintillation photon; it merely creates a new minority carrier and then a new photon in a random direction. The efficiency of photon collection in direct-gap semiconductors is therefore limited only by parasitic processes, such as nonradiative recombination of the minority carriers and free-carrier absorption of light. If these are minimized, one can have an opaque but luminescence transmitting semiconductor.

Both approaches have been studied using relatively thick (350 μm) S-doped n-type InP wafers. We studied the photoluminescence signal excited near the front surface of a wafer by a CW laser beam of λ = 640 nm. It has been shown that the ratio of spectrally integrated luminescence emerging from the back surface (transmission luminescence) increases with the carrier concentration $n$ from 7% at $n = 2 \cdot 10^{18}$ cm$^{-3}$ to 13% at $n = 8 \cdot 10^{18}$ cm$^{-3}$ at room temperature relative to the reflection luminescence emerging from the front surface. However, the total luminescence intensity drops with increasing $n$, which indicates the growing impact of non-radiative recombination (Fig. 1a). Our measurements of the luminescence decay time τ showed a $\tau \propto n^{-2}$ dependence for $n > 10^{18}$ cm$^{-3}$, which is characteristic of the Auger recombination [3]. At T = 77 K, the Auger recombination becomes negligible leading to an increase of the total luminescence intensity. The transmitted-to-reflected luminescence ratio also increases up to 45% for $n = 2 \cdot 10^{18}$ cm$^{-3}$ and up to 63% for $n = 8 \cdot 10^{18}$ cm$^{-3}$ as a consequence of the thermal shift of the absorption edge relative to the luminescence spectrum [3].

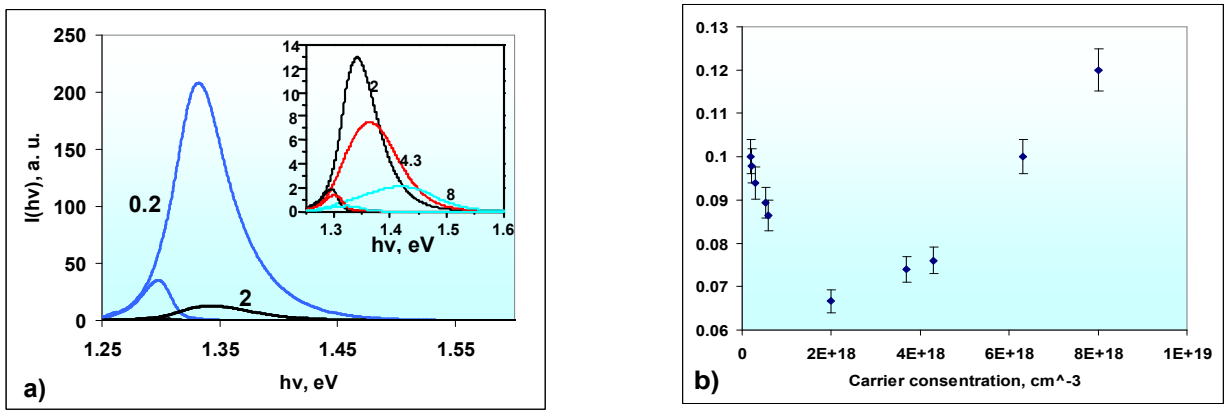

Fig. 1 (a) Front-side luminescence spectra and spectra emerged from the back surface of a wafer (smaller ones near the low-energy wings of the corresponding front-side spectra of the same color) for different carrier concentrations $n$ (in units of $10^{18}$ cm$^{-3}$) at room temperature; (b) Ratio of the integrated transmission luminescence to the reflection luminescence at room temperature for 640-nm excitation wavelength as a function of carrier concentration.

Surprisingly, for $n < 10^{18}$ cm$^{-3}$, the transmitted-to-reflected luminescence ratio starts growing again contrary to what could be expected from the absorption edge shift. At room temperature the ratio reaches 10% for $n = 2\cdot10^{17}$ cm$^{-3}$ (Fig. 1b). Moreover, at this low concentration the total luminescence intensity increases more than 10-fold in comparison with heavily doped samples (Fig. 1a). The observed increase of the luminescence intensity evidences a significant reduction in the nonradiative recombination rate while the rise in the transmitted-to-reflected luminescence ratio manifests an anomalous transparency of bulk InP to its own luminescence. This effect can be explained by a substantial enhancement of the photon recycling factor (Fig. 2 left), characterizing the average number of photon absorption/re-emission events per one initially created electron-hole pair. The enhancement of photon recycling at lower $n$ is owing to both lower free-carrier absorption and higher quantum radiative efficiency of luminescence $\beta = (1+ v_{nr}/v_r)^{-1}$, where $v_r$ and $v_{nr}$ are, respectively, the rates of recombination via radiative and non-radiative processes (also shown in Fig. 2 left).

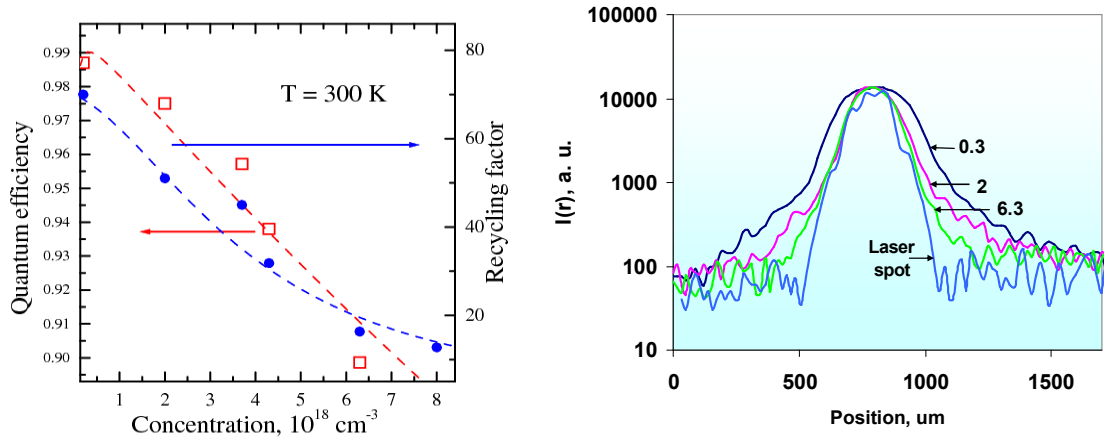

Fig. 2 (Left panel) Concentration dependence of the recycling factor $\Phi$ (squares) and quantum efficiency $\beta$ (dots) evaluated from the experimental data; dashed curves are analytical approximations [3]; (Right panel) Radial profile of the excitation laser spot (CW 930 nm) and distributions of luminescence around the excitation spot measured on the back sides of the wafers with various carrier concentration (in units of $10^{18}$ cm$^{-3}$).

The effects of anomalous propagation of photon-recycled luminescence through otherwise opaque direct-gap semiconductor and the corresponding photon-assisted diffusion of minority carriers are demonstrated in Fig. 2 (right). The presence of luminescence well beyond the laser spot indicates the existence of recombining holes seeded by the reabsorbed luminescence photons propagating radially from the excitation spot. Concentration of holes decreases with radius slower for the samples of lower concentration of majority carriers. The effect of anomalous diffusion was also observed earlier in InGaAs [4]. In a lossless situation with a fixed photon-assisted diffusion length, the distribution of minority carriers should be proportional to $\ln(r_0/r)$ in cylindrical coordinates, where $r_0$ is a radius of excitation spot. Analysis of this dependence delineates the extent of the lossless propagation of minority carriers and photons via the recycling effect.